\documentclass[3p,times,twocolumn]{elsarticle}
\usepackage{aas_macros}

\usepackage{ecrc}

\volume{00}

\firstpage{1}

\journalname{Nuclear and Particle Physics Proceedings}

\runauth{O. Tibolla \& R. Blandford}

\jid{nppp}

\jnltitlelogo{Nuclear and Particle Physics Proceedings}

\usepackage{amssymb}

\usepackage[figuresright]{rotating}

\begin{document}

\begin{frontmatter}



\dochead{}

\title{Cosmic Ray Origin $-$ Beyond the Standard Models}


\author{Omar Tibolla}

\ead{omar.tibolla@gmail.com}

\address{Mesoamerican Centre for Theoretical Physics (MCTP), Universidad Autonoma de Chiapas (UNACH), Carretera Emiliano Zapata Km. 4, Real del Bosque (Teran). 29050 Tuxtla Gutierrez, Chiapas, Mexico}

\author{Roger D. Blandford}

\ead{rdb3@stanford.edu}

\address{Kavli Institute for Particle Astrophysics and Cosmology, Department of Physics and SLAC National Accelerator Laboratory, Stanford University, Stanford, California 94305, USA}

\begin{abstract}

Given the success of the first meeting of ``Cosmic Ray Origin - Beyond the Standard Models'' (CRBTSM 2014), it was decided 
to hold a second meeting
of this international conference. 
In these 
introductory remarks, we 
rehearse the motivation for reconsidering the origin(s) of cosmic rays (CR).
We argue that the standard model, in which the majority of Galactic cosmic rays are produced through Diffusive Shock Acceleration (DSA) in SuperNova Remnants (SNR), is insufficient to account for recent observations. Some alternative scenarios are introduced and examined.
\end{abstract}

\begin{keyword}

Cosmic Ray Origin: Beyond the Standard Models, CRBTSM



\end{keyword}

\end{frontmatter}




\section{Introduction}

After the idea that the newly discovered anomalous ionisation might be of extraterrestial origin \cite{Wilson}, 
was proposed around 1900, several
pioneers begun to study the phenomenon in detail.
In particular,  Domenico Pacini  \cite{pacini} detected cosmic rays in Lake Bracciano and the Tirreno Sea in the years 1907-1912. In addition, Theodor Wulf \cite{wulf} \cite{wulf2}, a German Jesuit priest, discovered them on the Eiffel Tower in 1909 and introduced the description
``Hoehenstrahlung'' (``radiation from above''). Wulf's work was crucial from as technical standpoint because he invented the electroscope. In 1912, Viktor Hess used these electroscopes in his famous balloon flights and showed that the ionization was more intense at high altitude, work for which he shared the 1936 Nobel Prize.
Not everybody was fully convinced. The U.S. physicist Robert Andrews Millikan was continually skeptical about the conclusions of his European colleagues and disputed them for several years.
However he deserves to be mentioned because, ironically, he coined the modern description, ``cosmic rays''.
More details about the history of CR  can be found in \cite{crbtsm2014}.

The term ``cosmic rays'' generally includes all the energetic charged particles, with energies $E$  in the range from $\sim1\,{\rm MeV}\equiv10^{-3}\,{\rm GeV}$ to $\sim100\,{\rm EeV}\equiv10^{11}\,{\rm GeV}$, that reach Earth. 
The lowest energy particles, with $E\lesssim1$~GeV come mostly from the heliosphere and dominate the Galactic spectrum, which is therefore hard to measure at these energies. 
CR in the range $1\,{\rm GeV}\lesssim E\lesssim3\,{\rm PeV}\equiv3\times10^6\,{\rm GeV}$,  the ``ankle'', are 90\% protons and only $\sim$2\% electrons and positrons, despite the many bright, nonthermal sources powered by relativistic electrons that are observed. 
Their energy distribution as measured by satellites, is described by a power law with slope $-2.7$. Very High Energy (VHE) cosmic rays, with $3\,{\rm ~PeV}\lesssim E\lesssim3\,{\rm EeV}\equiv3\times10^9\,{\rm GeV}$, the ``ankle'', can be studied from the ground, using the Earth as a calorimeter. 
They have a steeper energy spectrum with slope $\sim3$ (up to $\sim3.3$ above $5 \times 10^{17}$ eV). The Ultra High Energy (UHE) CR with energies from the ankle to $\sim100$~EeV have a flatter slope that ends at the ``GZK'' cutoff caused by photopion production on the cosmic microwave background.  

\section{Standard Models and Beyond}

CR have long been associated with SNR
\cite{baade}. The  ``standard model'' of their origin derives from the monograph \cite{ginzburg} and its modern version  is described in articles in  \cite{crbtsm2014} and reviews cited in \cite{roger2014}. According to this model, primary Galactic CR up to the ``knee'' are accelerated in SNR shells which form at a rate of $\sim1$ per 50 yr per galaxy.  
In order to account for the CRs energy density of $\sim$1 eV$/$cm$^3$ and their confinement time deduced from CRs spallation, the typical CR energy release per supernova has to be $\sim10^{50}$~erg, which is about 10\% of the kinetic energy released in SN explosions
This idea was supported at the end of the 80's by the fact that this energy yield prediction of the ``standard model'' agreed with observations of the acceleration of relativistic particles in SNR shocks \cite{heinz} \cite{luke} \cite{luke2}. An additional factor was the developement, in the late seventies, of a natural way to accelerate CRs in SNRs shells, a variant of the Fermi acceleration mechanism \cite{fermi} associated with strong shock fronts: the Diffusive Shock Acceleration (DSA) theory \cite{dsa1} \cite{dsa2}\cite{dsa3} \cite{dsa4}. See \cite{crbtsm2014} for more details of the modern version of this theory. 

This standard picture of CR origin up to the ``knee'' was strengthened by the detection of TeV gamma-rays from SNRs spatially coincident with the sites of non-thermal X-ray emission by Imaging Atmospheric Cherenkov Telescopes (IACTs) (e.g. \cite{velajr} \cite{1713}). 
In addition, \emph{Fermi}-LAT and \emph{Chandra} observations of prominent SNR such as Tycho confirmed that the particle acceleration was mostly associated with the protons, accounting for $\sim10-20$\% of the explosion energy (e.g. \cite{slane}, \cite{damiano}.  Further support came from detailed studies of chemical and isotopic CRs composition \cite{ellison}.

However, serious problems also arose.  The TeV gamma-ray spectra of SNR shells seem to cut off at lower energies than predicted. \emph{Fermi}-LAT observations of the prominent SNR RX J1713.7-3946 seemed to favor leptonic, not hadronic, acceleration
e.g. \cite{heinz17},
\cite{lat1713}.
In another well-studied, young SNR,  Cassiopeia A, the particle acceleration fraction of the kinetic energy was measured to be $\sim$2\%.
\cite{casa} \cite{casb}.
The response to these shortcoming of the standard model of Galactic CR has been either to modify it or to replace it.

There has also been a growing interest in understanding the origin of the UHECR. On quite general grounds the sources have to be associated with large luminosity, typically $\gtrsim10^{45}\,{\rm erg s}^{-1}$ for the highest energy particles before they lose energy. This cuts down the number of choices. They include Gamma  Ray Bursts, relativistic jets associated with active galactic nuclei, rapidly spinning magnetars and shocks in the intergalactic medium especially associated with infall onto rich clusters of galaxies.

This was the context for  ``Cosmic Ray Origin: Beyond the Standard Models'' (CRBTSM), a dedicated international conference, held in San Vito di Cadore, one of Enrico Fermi's favoured holiday resorts (e.g. \cite{sanvitofermi}).

\section{CRBTSM  2014}

We organized the first CRBTSM meeting in March 2014 around five questions:

\begin{itemize}

\item{What evidence do we have for the SNR origin?}

\item{What other sources might there be in the Galaxy?}

\item{What causes the knee?}

\item{Where (in energy) is the transition to an extra-Galactic component?}

\item{What extra-Galactic sources are conceivable?}

\end{itemize}

The summary of our discussions of these questions was captured in the five chapters of the CRBTSM 2014 monograph book \cite{crbtsm2014}.
We began with the strong evidence in support to the standard model.
Measured primary and secondary CR nuclei spectra as well as the positron-electron ratio are consistent with the Galactic CR origin in SNR up to $\sim 10^{17}$ eV \cite{evgeny}.
The observations of SN 1006 \cite{evgeny} \cite{fra} and in particular of the the above-mentioned Tycho SNRs \cite{evgeny} \cite{fra} \cite{caragiulo} give strong support to the hadronic scenario and the standard model.
Although their interpretation is more complicated,  SNR interacting with Molecular Clouds (MC) also appear to support the standard model, as shown in the case of the W44, W28 and IC443 SNR/MC systems \cite{martina} \cite{pi0}. 

The association of CR and SNR is still the crux of the argument in favor of the standard model \cite{pasquale} \cite{damiano2}.
However the growth of the CR-induced streaming instability seem barely adequate to account for energies of $\sim 3Z\,{\rm PeV}$ despite the advantage conferred by field amplification at SNR shock fronts and an appeal to very energetic SNe in red giant progenitor stars winds \cite{pasquale}.
Nevertheless, if SNR are responsible for proton acceleration up to the knee in the spectrum, (1) there should be a gradual transition to heavier elements and (2) Galactic CRs should cease at $\sim100\,{\rm PeV}$ \cite{pasquale}, in good agreement with the observations \cite{evgeny}.
The tentative association of very thin optical filaments dominated by Balmer $H_{\alpha}$ emission with SNR shocks seems to confirm an efficient CR acceleration at SNR shells \cite{morlino}.
Additional features of the observations can be attributed to CR propagation across the Galaxy \cite{zirak}.

Alternative CR accelerators may also contribute to the observed Galactic spectrum. Examples include massive binary systems in open clusters \cite{bednarek}, protostellar jets
and runaway stars \cite{anabella}, the Fermi Bubbles, which are also suggested to be one of the possible causes of the spectral feature at the \emph{knee} \cite{cherny}, the Galactic center itself \cite{thoudam}, pulsars \cite{kotera}, Pulsar Wind Nebulae \cite{weinstein} \cite{crbtsm2014}. 
In addition,  extragalactic sources, such as gamma-ray bursts \cite{meszaros}, galactic shocks \cite{meszaros}, starbursts galaxies \cite{yoel}
and active galactic nuclei \cite{karl}, can also play a role, especially at higher energy.
Extragalactic sources are generally supposed to be responsible for CR acceleration above the ankle, though some Galactic sources, such as rapidly spinning magnetars, have been proposed for the acceleraton of UHECR.

Our understanding of CR acceleration has been greatly influenced by indirect, electromagnetic  observations of putative sources. In particular, gamma-ray satellites such as Ferrmi and AGILE together with atmospheric Cerenkov observatories have made many important contributions, studying injection, mechanisms and escape of CRs. Future X-ray telescopes will teach us much more about SNR and similar sources \cite{aya}.
A large fraction of CRBTSM 2014 was devoted to the studies of the \emph{knee} region and above, i.e. to other experiments, which are crucial for CRs studies, and to the interpretation of their data. 
Direct and indirect CRs observatories were discussed, such as: KASCADE-Grande \cite{bertaina}, PAMELA \cite{emi} \cite{panico}, BESS \cite{emi}, AMS-02 \cite{emi}, Tunka-133 \cite{lyubov}, CALET \cite{calet},
ATIC \cite{panov}, Pierre Auger Observatory \cite{etienne} and the future JEM-EUSO \cite{bertaina2}.
All these facilities are allowing us to explore beyond the SNR-CR association to investigate alternative sources both Galactic and extragalactic \cite{etienne}. 
The future importance of neutrino astronomy was also emphasized \cite{bednarek} \cite{marcowith} \cite{fargion} \cite{karl}. Many techniques have been developed including water and ice detectors \cite{marcowith} \cite{karl} together with a new design, located in a deep valley \cite{fargion}.

Lengthy consideration of these topics contributed to the success of CRBTSM 2014 and motivated us to organize CRBTSM 2016 in the intimate and inspiring locale of San Vito di Cadore.


\section{CRBTSM 2016}

Two factors motivated our reconvening CRBTSM, the timeliness and general importance of the five questions that were used to organize CRBTSM 2014 and the progress that was being made in addressing them. It was decided to follow success and adopt a similar format that encouraged provocative plenary talks, allowed ample time for free-wheeling general discussion and scheduled long breaks for more specialized investigations. The meeting was judged by the participants to have been even more successful than its predecessor. Highlights and examples from the meeting include:
\begin{itemize}
\item{The propositions that all cosmic rays are Galactic and that they are all extragalactic were critically examined. The standard arguments that they are mostly Galactic below the knee, mostly extragalactic above the ankle and of mixed though less certain provenance in between were sharpened to the satisfaction of most though not all participants. (See contributions by Eichler and Strong. A very important participant noted that ``On Monday we were told that all CRs are Galactic. On Tuesday, we were taught that they all have an extragalactic origin. Since Wednesday, I have been confused!'').}
\item{The number of distinct components contributing to the total cosmic ray spectrum was debated. (See contribution by Tilav.)}
\item{The increased prominence of numerical simulations especially of global cosmic ray propagation was quite noticeable. These supported the general conclusions from simpler and less detailed analytical calculations and furnished testable predictions for the increasingly accurate observations of cosmic ray spectra and composition that are being reported. A notable and prescriptive example is the Boron to Carbon ratio. (See contribution by Blasi.)}
\item{There is increasing attention paid to the physical description of cosmic ray propagation, in particular to replacing spatial diffusion by a more sophisticated local description based upon kinetic simulations. (See controibutions by Caprioli and Cerutti.)}
\end{itemize} 

All CRBTSM 2016 talks are available in the official web-page: http://crbtsm.eu.





\section*{Acknowledgements}

This conference was supported by the 'Helmholtz Alliance for Astroparticle Physics HAP' funded by the Initiative and Networking Fund of Helmholtz Association.
O.T. acknowledges the grants Conacyt CB-258865, Conacyt-281653 and Royal Society NAF-180385; O.T. thanks Sarah Kaufmann and Massimo Persic for the support and for the useful discussions.
We would like to warmly thank all the people who contributed to the success of the second edition of CRBTSM: the SOC, the LOC (notably the secretary Sandra Calore),
the participants (most of whom have contributed to this book). Giovanni Bignami gave a public lecture concluding CRBTSM 2016 activities. With colleagues all around the world, we mourn his passing and salute his remarkable contributions to particle and high energy astrophysics. 
We also extend our heartfelt thanks to the San Vito di Cadore town administration (notably Mayor Franco De Bon and vice-Mayor Andrea Fiori) for their warm hospitality and support, the 'consorzio Dolomiti' (notably its director, Massimiliano Forgiarini, and Bruno Talamini) for their constant support and Loris Felici for local technical support and for his ready availability to solve urgent problems.

\nocite{*}
\bibliographystyle{elsarticle-num}
\bibliography{martin}










\end{document}